\documentclass[11pt]{article}
\pdfoutput=1
\usepackage{jcappub}
\usepackage{natbib}
\usepackage{amsmath}
\usepackage{color,colordvi}
\usepackage{booktabs, multirow}
\usepackage[mediumspace,Gray,squaren]{SIunits}
\usepackage{hyperref}
\usepackage{aas_macros}
\usepackage[utf8]{inputenc}
\usepackage{amssymb,amsfonts}
\definecolor{darkblue}{rgb}{0,0,0.3}
\definecolor{darkgreen}{rgb}{0,0.3,0}
\hypersetup{
  pdftitle = {Machine Learning Etudes in Astrophysics},
  pdfauthor = {Hajian, Alvarez, Bond},
  colorlinks,%
  citecolor=darkblue,%
  filecolor=black,%
  linkcolor=blue,%
  urlcolor=blue
}

\definecolor{orange}{rgb}{1,0.5,0}

\addunit{\jansky}{\ensuremath{\mathrm{Jy}}}
\addunit{\degword}{\ensuremath{\mathrm{deg}}}
\newcommand{\be}{\begin{equation}}
\newcommand{\ee}{\end{equation}}
\newcommand{\bea}{\begin{eqnarray}}
\newcommand{\eea}{\end{eqnarray}}
\newcommand{\rmn}{\mathrm}

\begin{document}

\title{\huge{Machine Learning Etudes in Astrophysics:}\\ \LARGE{Selection Functions for  Mock Cluster Catalogs}}

\author[a]{\Large{Amir Hajian,}}
\emailAdd{ahajian@cita.utoronto.ca}
\author[a]{Marcelo Alvarez,}
\author[a]{J.~Richard~Bond}

\affiliation[a]{Canadian Institute for Theoretical Astrophysics, University of Toronto,\hspace{1em}Toronto, ON M5S~3H8, Canada}

\abstract{Making mock simulated catalogs is an important component of astrophysical data analysis. Selection criteria for observed astronomical objects are often too complicated to be derived from first principles. However the existence of an observed group of objects is a well-suited problem for machine learning classification. In this paper we use one-class classifiers to learn the properties of an observed catalog of clusters of galaxies from ROSAT and to pick clusters from mock simulations that resemble the observed ROSAT catalog. We show how this method can be used to study the cross-correlations of thermal Sunya'ev-Zeldovich signals with number density maps of X-ray selected cluster catalogs. The method reduces the bias due to hand-tuning the selection function and is readily scalable to large catalogs with a high-dimensional space of astrophysical features.}

\maketitle
\flushbottom

\section{Introduction}
While studying astrophysical objects we often come across problems where a thorough analytical calculation of the underlying physical theory is formidable. These problems are usually solved by numerical methods including large simulations that contain realistic and complicated physical processes of the system. Such simulations typically generate much more data than  actually needed to solve our problem. For example, if we are only  interested in  clusters of galaxies in a specific mass range and we use mock simulations to generate a catalog of dark matter halos, we will obtain a large sample of all kinds of clusters that we should trim to procure what we need for our purpose. The selection criterion used in putting together a sub-sample is called the \textit{selection function}. If a selection function is known, picking objects from simulations that have physical properties determined by the selection function is trivial. But determining the selection function when dealing with catalogs derived from combining various astrophysical observations is a challenging task. In many cases the number of unknown parameters is so large that explicit rules for deriving the selection function do not exist. A sample of the objects does exist (the very objects in the observed catalog) however, and the observed sample can be used to  express the rules for the selection function. This ``learning from examples'' is the main idea behind classification algorithms in machine learning. The problem of selection functions can be re-stated in the statistical machine learning language as: given a set of samples, we would like to detect the soft boundary of that set so as to classify new points as belonging to that set or not.

In this paper we revisit the problem of making a catalog of simulated dark-matter halos. We would like to make a synthetic catalog to  resemble a galaxy cluster sample selected based on their X-ray properties. The X-ray sample is the same as the one used in our earlier paper  to measure the thermal Sunya’ev-Zeldovich power spectrum using cross-correlations of Planck and ROSAT \cite{hajian2013measuring}. 
This simple example is used to demonstrate the application of machine-learning in automating a widely used statistical procedure in analyzing the astrophysical data. 
 Our approach is to make a description of a target set of observed clusters \footnote{Clusters in this paper refer to galaxy clusters, not to the clusters one often searches for in data mining.}
 of galaxies and to detect which  clusters in a new data-set (\textit{e.g.} a simulated catalog) resemble this training set. All other clusters are outliers by definition and need to be eliminated from the catalog. 
The strength of this approach becomes more evident as the dimensions of the  feature space (\textit{i.e.} the number of astrophysical properties by which we select and identify the catalog) increases. However to be able to compare it with an already solved problem we stick to the X-ray selected cluster catalog used in \cite{hajian2013measuring} as an example of an observed sample and demonstrate the generation of a list of objects from a large set of simulated dark matter halos in a mock simulation\footnote{Mocks in general may be based on simulations using gas-dynamical, N-body or semi-analytic techniques. Here we use the peak-patch picture which is a semi-analytic method of generating mock halos (see Section \ref{sec:sim}). But the methods described here can be applied to any mock simulation.}, so that the picked objects look like the objects in the observed catalog in every astrophysical aspect. Wherever needed, we present a concise introduction to the machine learning concepts. For a more detailed explanation of the terminology see \textit{e.g.} \cite{bishop} and \cite{astroML}. 
 Machine learning techniques have been used in various astrophysical applications in the (recent) past, such as to classify survey images and galaxy morphologies, to detect variables and transients in large surveys, to measure photometric redshifts more accurately, etc. Support vector machines was used by \cite{xu2013first} to populate dark matter halos with galaxies.  For a list of applications of machine learning in astrophysics and references to the associated extensive literature see the reviews in the books  \cite{MLadvances, astroML}.

This paper is organized as follows: 
Section \ref{sec:meth} describes one-class classifiers with a focus on the two methods used in our analysis.  Section \ref{sec:results} explains the data-sets and simulations used in this work. Section \ref{sec:results} also outlines the practical application of the algorithms to making mock catalogs of clusters of galaxies and measuring their statistical properties. We draw our conclusion in Section \ref{sec:concl} and present some details of mass conversions in the Appendix \ref{app:mass}.

\section{One-Class Classification} \label{sec:meth}
In most classification problems, the goal is to classify data into two or more classes based on common patterns in each class. It is often assumed that data from all classes are statistically balanced, \textit{i.e.} equally sampled. One-class classification (used in this analysis) is a severe departure from this case, in which only one class of data is available. 
The goal is then to distinguish between objects from one class,   the \textit{target class}, and all other possible objects,  \textit{outliers}, assuming that only  examples of the target class are available to us to train our algorithm. 
The target class is the observed sample in hand. Therefore only information of the target class  (and not the outlier class) is available to us through measurements. Our task here is to define a boundary around the target class to accept as many of the target objects as possible, and to minimize the chance of accepting outlier objects.
Two main approaches to one-class classification used in this paper are\footnote{See \cite{tax2001thesis} for a review of one-class classifiers and their properties. }:
\begin{itemize}
\item Density Estimation Methods:  This method is widely used in astrophysics.   It is based on estimating the density of the training data and setting a threshold on this density to define the boundary of the class. A commonly used shape for the density is the Gaussian distribution function (or a linear combination of a number of them). This method is simple. It works well when a reasonably large sample size is available to allow for a proper estimation of the underlying density. 

\item Boundary Determination Methods: 
When only a limited amount of data is available, the feature space\footnote{The feature space is the space of the properties used in classification. Here our feature space is the two-dimensional $M-z$ (mass-redshift) space. } is not well-sampled and the distribution function of the data will be poorly constrained. Hence a full description of the data density might be too demanding when we only need to determine a boundary for classification. This method computes a binary function that captures regions in the input space where the probability density of the data lives, such that most of the data lie in the region where the function is non-zero \cite{scholkopf1999support}.  One-class SVM is an example of this type of method that we use in this paper.
\end{itemize}

\begin{figure}[t!]
\centering
\includegraphics[scale=0.4]{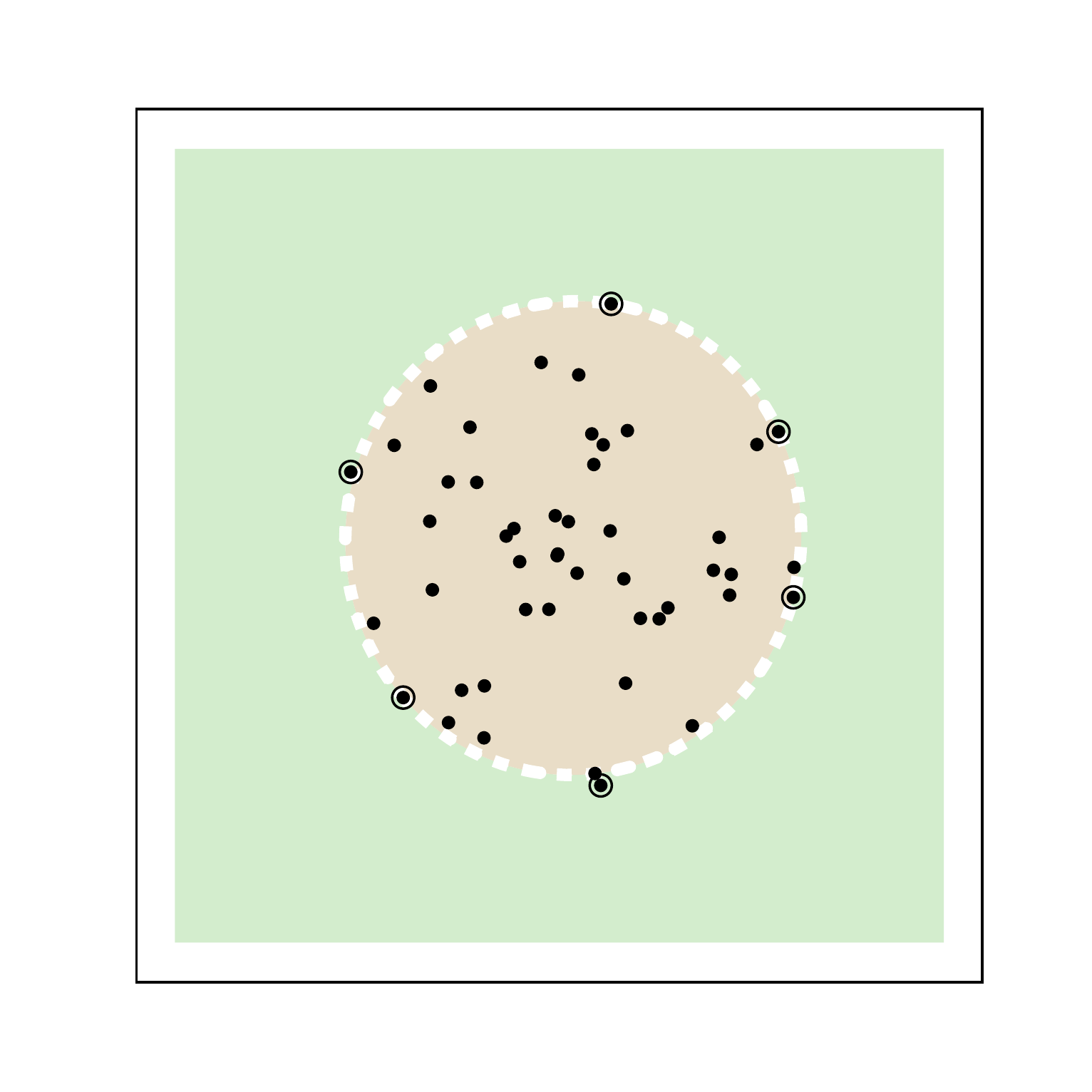}
\includegraphics[scale=0.4]{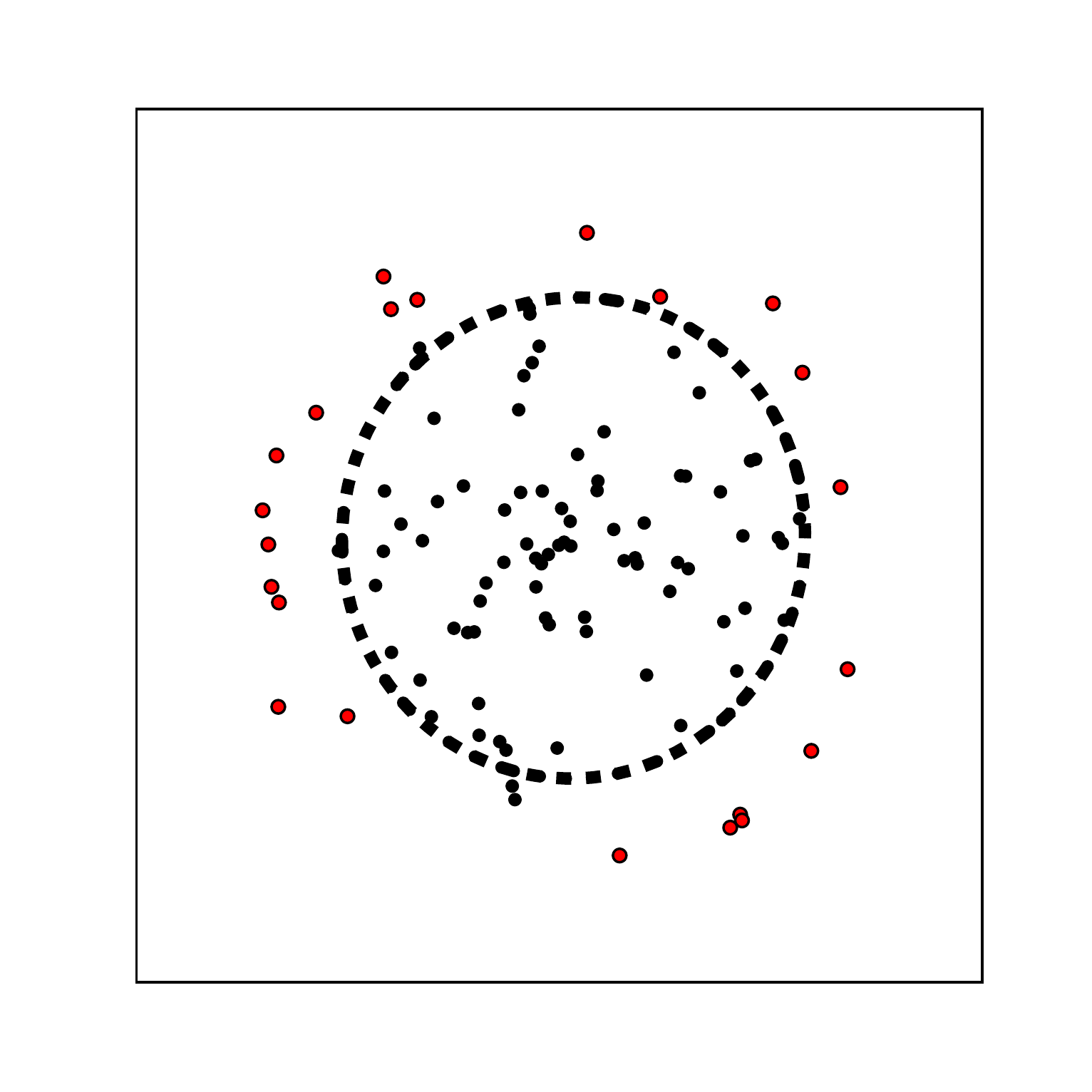}
\caption{Example of a One-Class SVM Classifier. \textit{Left:} unsupervised learning of the  decision boundary (dashed white surface) on a training set. Samples on the margin are called the support vectors. \textit{Right:} supervised classification of new data based on the learned decision surface. Color coding shows the actual classes of the simulated data: black data points are members of the target class, while red points are outliers by construction.}
\label{fig:1class}
\end{figure}

\subsection{Support Vector Machines for One Class Classification} 
\label{sec:svm}
Support vector machines (SVMs) are a family of supervised learning methods used in a wide range of applications including classification, regression and outlier detection. The SVMs are effective in high dimensional spaces even when the number of samples is less (but not much less) than the number of dimensions. The most important property of the SVMs is their versatility through the use of kernel functions adaptively designed for specific purposes. 
A one-class support vector machine is an unsupervised algorithm  that learns a decision function for outlier  detection, \textit{i.e.} classifying new data as similar or different to the training set. This algorithm attempts to learn a decision boundary to separate the region containing the observed samples from the rest of the space \cite{scholkopf1999support, tax2004support}.

Mathematical formulation of one-class SVM is simple. Consider a dataset with $N$ objects, 
$D = \{\mathbf{x}_1, \mathbf{x}_2, \cdots, \mathbf{x}_N \}$. Each $\mathbf{x}_i$ represents a feature vector in $d$-dimensions for training examples belonging to one class $X$. We desire a decision function $f(\mathbf{x})$ whose sign determines the target region and separates it from the outliers. In order to make the SVM work in linearly inseparable cases (\textit{i.e.} where the classes can not be separated by a linear boundary surface), the original feature space needs to be mapped into a higher dimensional space in which the classes are separable. This can be costly and a \textit{kernel trick} is often used to keep the computational cost manageable \cite{scholkopf2002learning,cristianini2000introduction}. The use of kernels removes the complexity of introducing higher dimensional spaces and keeps the complexity at the feature space dimension level.   The exact form of the kernel function is chosen based on the problem. Some popular kernel functions often used in the SVM method are
kernels based on dot products, \textit{e.g. } the polynomial kernel, $\left(\gamma \mathbf{x}_i \cdot \mathbf{x}_j + r\right)^d$, and
distance-based kernels, \textit{e.g.} the Gaussian radial basis function (RBF), $k(\mathbf{x}_i, \mathbf{x}_j) = \exp\left[\frac{|\mathbf{x}_i- \mathbf{x}_j|^2}{\gamma^2}\right]$.  The free parameters  in kernel functions (such as $\gamma$, $d$ and $r$) are determined by cross-validation as described in Section \ref{sec:training}.

\begin{figure}[t!]
\centering
\includegraphics[scale=0.6]{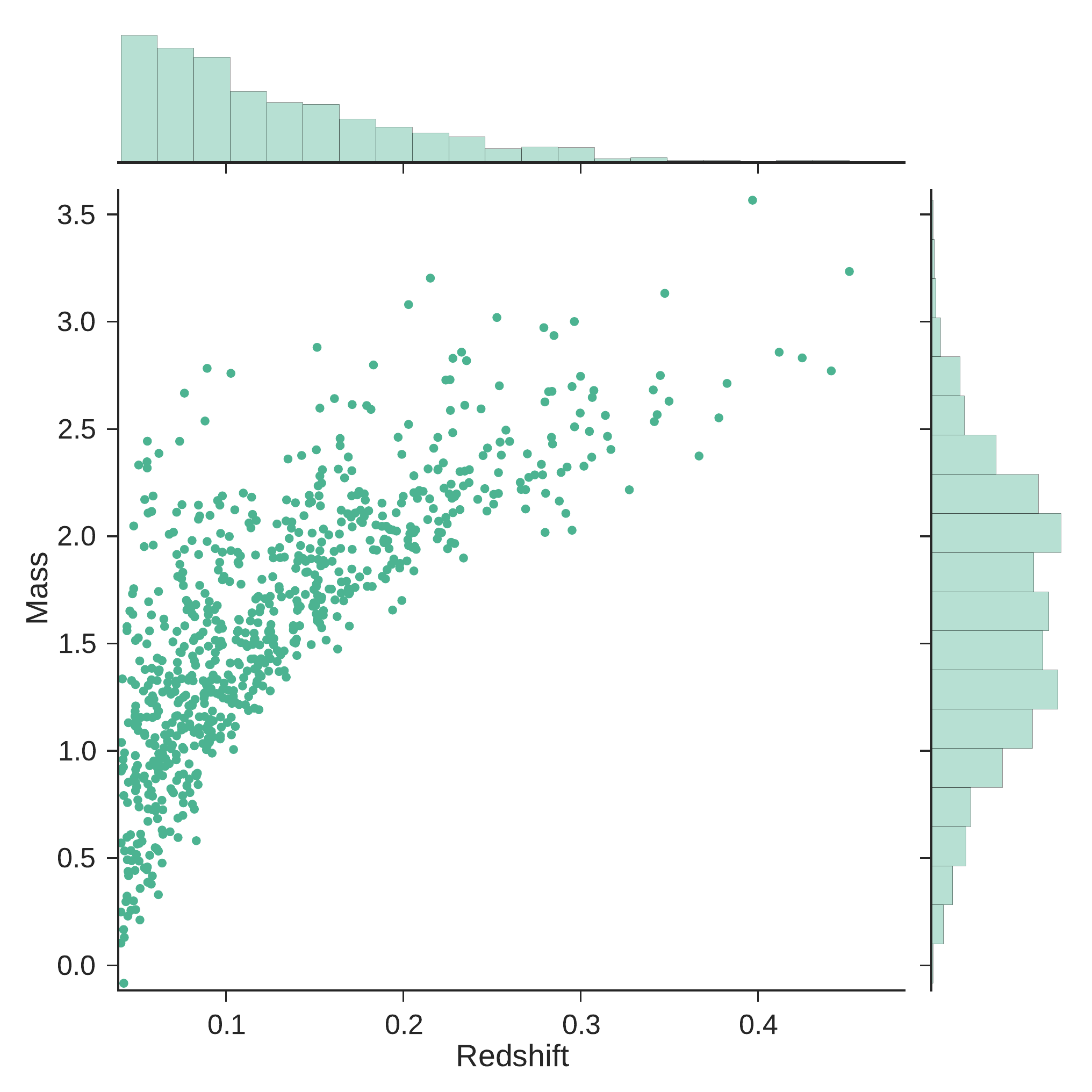}
\caption{Distribution of our galaxy cluster sample in the Mass-Redshift space. The mass on the $y$-axis is the rescaled mass,  $\log{M_{vir}}[10^{14}M_\odot]$.}
\label{fig:jointPlot}
\end{figure}


Our goal is to find a decision function, a binary function that returns $+1$ in a minimal region containing the training data points and $-1$ elsewhere. It can be computed by finding a hypersurface in the higher-dimensional space that separates the data in the training set (the $+1$ region) from zero (the boundary of the $+1$ and $-1$ regions). That translates into finding a small region that contains the training set 
in the mapped space. To that end one needs to solve the following dual optimization problem using quadratic programming\footnote{In mathematical optimization, quadratic programming is  the problem of finding the maximum or minimum of a quadratic function of several variables, possibly subject to linear constraints \cite{wright1999numerical}.}:
\bea \label{eqn:optimization}
\min_{\alpha}\frac{\alpha^T K \alpha}{2}& & \\ \nonumber
\mathrm{subject\,\, to\,\,} 0 \leq \alpha_i \leq \frac{1}{\nu N}&,&\sum_i{\alpha_i}=1,
\eea
where $\alpha$'s are the weights that need to be learned through solving the above problem, $K_{ij}=k(\mathbf{x}_i, \mathbf{x}_j)$ is the kernel matrix and $\nu \in (0,1)$ is a parameter that characterizes the solution by 
setting an upper bound on the fraction of outliers, \textit{i.e.} training examples regarded as not belonging to the target set. Therefore $\nu$ is a lower bound on the fraction of training examples used as support vectors, \textit{i.e.} the subset of the training data that lie on the margin.  The classification task in the SVM is only a function of the support vectors.

The decision function is then uniquely given by the solution to the above equations:
\begin{equation}
f(\mathbf{x}) = \mathrm{sgn}\left(\sum_{i=1}^{N}{\alpha_i k(\mathbf{x}_i, \mathbf{x}) - \rho} \right),
\end{equation}
in which $\rho$ is the offset determined by the optimization solution (see \textit{e.g.} \cite{scholkopf1999support}).

We demonstrate the One-Class SVM using a simple example in Figure \ref{fig:1class}.  The training sample is a dataset drawn from a radially uniform random distribution between 0 and $R$. Using an RBF Gaussian kernel we compute the decision surface that encloses the target class and separates it from the rest of the space. The boundary is effectively defined by the outermost samples (the \textit{support vectors}), a fraction of which is considered outliers. Setting the outlier fraction depends on the details of the problem and is defined by the physical properties and uncertainties in the training sample.  This free parameter is set by $\nu$ as described in Section \ref{sec:svm}. The left panel of Figure \ref{fig:1class} shows the learned decision boundary.   The right panel shows the result of applying the learned decision function to a new dataset in order to classify it into a target class that looks like our training sample (data points inside the decision boundary) and outliers (everything else). The new data in the right panel of Figure \ref{fig:1class} were generated using a population drawn from the same distribution as the training sample (black data points) and another population drawn from a wider distribution function (red data points). The  decision function trained on the training data of the left panel successfully separates the two populations of the new data in the right panel of Figure \ref{fig:1class}. This is a simple two-dimensional example and the decision function of Figure \ref{fig:1class} can be computed without a machine-learning method. But increasing the dimensions of the feature space even by one, or modifying the distribution function to be non-symmetric, will quickly make the simple, hand-drawn methods inaccurate (if not impossible).

\subsection{Gaussian Mixture Model}

\begin{figure}[t!]
\centering
\includegraphics[scale=0.5]{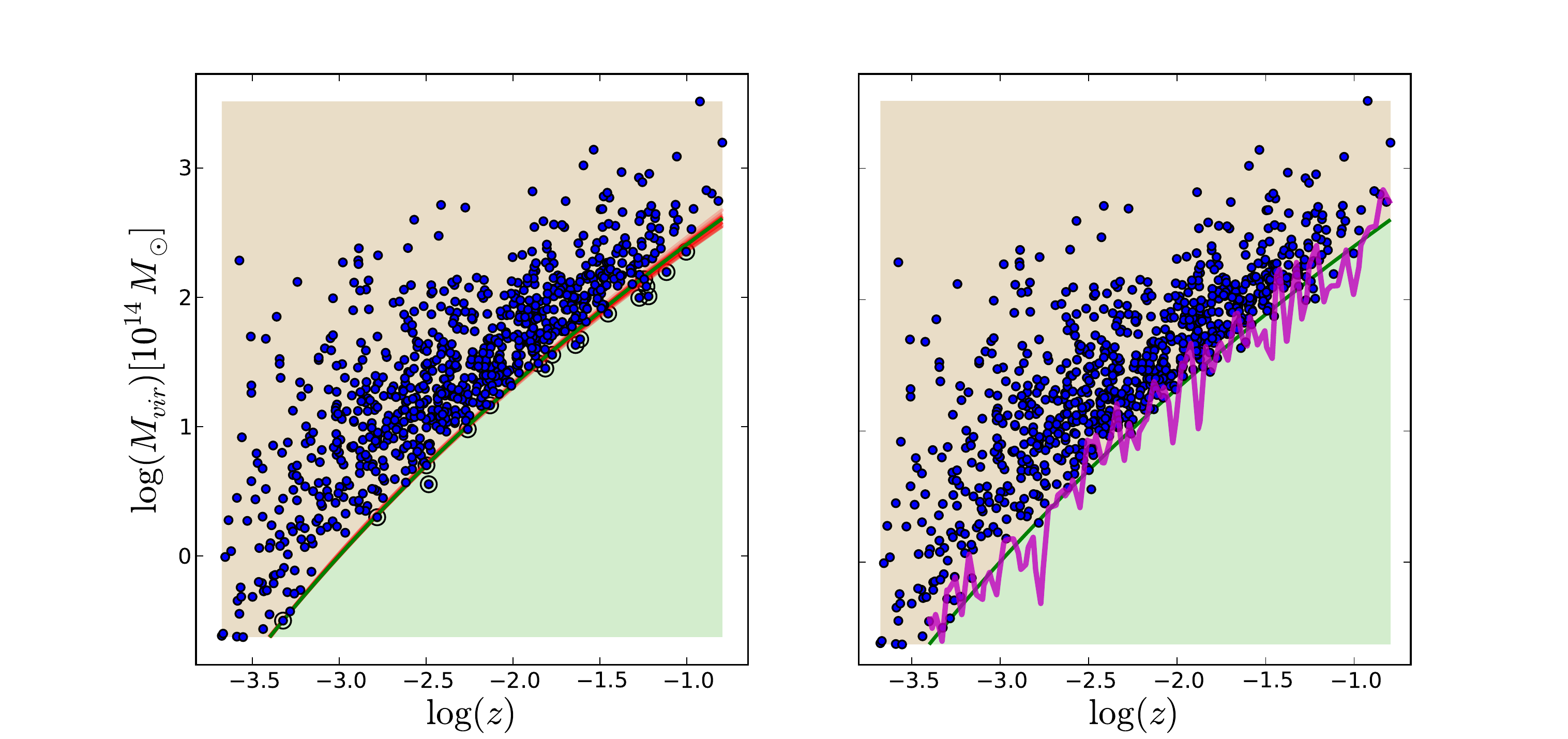}
\caption{Learned decision boundary for the galaxy cluster sample used in our analysis. The decision surface separates the $M-z$ plane into the target space (above the surface) and the outlier space (below the line). Clusters in the target space belong to the same class as the observed X-ray cluster catalog, while the outliers won't be observable. (\textit{Left:}) the red lines are decision surfaces at each cross-validation realization. The green line is the decision boundary obtained from using all clusters in the sample. Support vectors are shown by a different point style. (\textit{Right:}) The sharp decision boundary is softened by introducing the uncertainties into the model. The magneta surface is a realization of a decision boundary used in classification.}
\label{fig:learnedSurface}
\end{figure}

A \textit{Gaussian Mixture Model} (GMM) is a linear combination of $K$ normal distributions of the form
\begin{equation}
p(x) = \sum_{i=1}^{K}{\alpha_i \mathcal{N}(\mathrm{x}|\mu_i,\Sigma_i)},
\end{equation}
where each \textit{component}, $\mathcal{N}(\mathrm{x}|\mu_i,\Sigma_i)$, is a Gaussian distribution function with mean $\mu_i$ and covariance $\Sigma_i$ \cite{bishop}. Mixing coefficients, $\alpha_i$, are non-negative
\begin{equation}
{\alpha_i} \geq 0, \forall i
\end{equation}
and  satisfy the completeness condition
\begin{equation}
\sum_{i=1}^{K}{\alpha_i}=1,
\end{equation}
the latter to normalize $p(x)$, 
\begin{equation}
0 \leq \alpha_i \leq 1, \forall i. 
\end{equation}
In practice, the number of mixture components are defined beforehand, and the means and covariances are estimated from the data. Cross-validation\footnote{See Section \ref{sec:training} for a description of the cross-validation step.} is used to determine the number of clusters in our study. We use \texttt{scikit-learn},  an open source machine learning library for Python \cite{scikit-learn}, for the implementation of the one-class SVM, Gaussian mixture models and cross-validation in our analysis.

\section{Application to X-ray selected clusters}\label{sec:results}
In what follows we use a one-class SVM to select all dark-matter halos in a mock  simulation whose sub-components resemble an observed cluster catalog. This is a large catalog that contains all halos in the simulation that belong to the same class as the observed sample but may have orders of magnitude larger number of objects than the observed sample. We call it the super-catalog and in making it we make no assumptions beyond the detectability of the clusters.  This is a statistical alternative to estimating the ``observation selection function'' by hand and using it as a threshold for making the catalog. We also train a generative GMM classifier to sub-sample that super-catalog  to make a simulated cluster sample that has the same distribution function in mass and redshift as the observed sample. 
\subsection{Cluster Catalog}
We use the X-ray selected cluster catalog of \cite{hajian2013measuring} that is a sub-sample of the MCXC X-ray cluster catalog \citep{piff2011}. The MCXC sample combines the REFLEX\citep{bohr2004}, BCS\citep{ebel1998,ebel2000} and CIZA\citep{ebel2002,koce2007} flux limited catalogs using the RoSAT all sky survey \citep[RASS;][]{voge1999}. We cut the catalog at $z = 0.04$ as in \cite{hajian2013measuring} to reduce the cosmic variance from low redshift rare objects. The subsample chosen this way has 820 clusters of galaxies in it. The cluster masses were taken from the MCXC catalog, which uses a $L_\rmn{x}-M$ scaling relation to measure the $M_{500}$ mass of the clusters. We follow the procedure in \cite{mody2012one} to convert the $M_{500}$ masses in the MCXC catalog to virial mass (see Appendix \ref{app:mass} for more details).

Figure \ref{fig:jointPlot} shows our cluster catalog as a function of mass and redshift. 

\subsection{Simulation} \label{sec:sim}
We have recently developed a massively parallel implementation \cite{pykp} of the peak-patch algorithm for fast structure formation simulations \cite{pkp1}. The key point is that halos form via gravitational collapse of peaks in the matter distribution \cite{pkp1, pkp2, pkp3}. If the shape of a peak and its surrounding tidal field are measured, one can predict the position, mass, and time at collapse very accurately by using adaptive filtering and applying ellipsoidal collapse to each peak individually, based solely on the {\em initial linear density field}. The resulting halo catalogs  match closely those of much more expensive cosmological N-body simulations.

Given a halo catalog, efficient Monte Carlo construction of three-dimensional catalogs of objects such as groups and clusters of galaxies, and, with more selection criteria, of galaxies in their formation phases, is possible.
Of course, the results presented here are independent of the specific method used for generating the simulated halo catalogs, whether via N-body, gasdynamical, or other semi-analytic simulation methods. 

We use a peak-patch light cone run out to $z\sim 1.04$  and down to $M_{halo} \sim 1.5\times 10^{13} M_\odot$. It takes about 90 minutes on 250 cores to run for a volume of $4^3$ Gpc$^3$ and roughly yields a 2 Gpc sphere. The effective resolution of our dark-matter halo simulation is equivalent to an N-body simulation with $2560^3$ particles. The halo catalog used in this paper contains about $10^7$ objects. Details of generating the halo catalog are discussed in an accompanying paper \cite{pykp}.

\subsection{Training}\label{sec:training}
We first train a one-class SVM classifier with a third order polynomial kernel on our cluster catalog to compute the decsion function that separates the $M-z$ space into two parts: an observable region (target class) and an outlier region. Figure \ref{fig:learnedSurface} shows the data and the decision boundary. Almost all objects are above the decision boundary. This result is intuitive: objects on the upper left corner of the plot are nearby massive objects, therefore easily observable. Objects below the decision boundary are small and far, hence hard to observe. The choice of a
boundary that selects points above a redshift dependent threshold  
is motivated by the fact that a physically meaningful decision surface should not be bounded from above. Because objects at the top of our training sample are massive, they should be relatively easily detectable and hence should always belong to the target class.  We validate our decision function by cross-validation: we randomly choose 20\% of the objects as the test sample by shuffling the training sample and drawing a sub-sample containing  20\% of the data. The rest of the objects are used to train the algorithm and find the decision function. We then use the learned decision boundary to score the test sample, \textit{i.e.} measure the number of outliers in the test sample (called the rejection score). This procedure is repeated 25 times to yield an average score. We choose a setup that keeps the rejection score at about 25\%. This is achieved at $\nu=0.3$. This free parameter, as explained in Section \ref{sec:svm}, determines the fraction of the objects that are believed to be outliers contaminating the data. Its choice depends on the specific problem.  
Our choice is motivated by the average uncertainty in the masses of the clusters. 
Varying this parameter changes the  boundary surface in a predictive way: the smaller the outlier fraction, the lower the threshold in each redshift, and hence the more low-mass objects in the final sample. The effect of this parameter is diluted by how we soften the hard boundary of the SVM into a fuzzy surface. 
The boundary surface derived this way isolates the most deviant objects.  Figure \ref{fig:learnedSurface} shows the 25 learned decision boundaries in red and the decision boundary obtained by using the full dataset in green. We use this family of decision boundaries to draw realization samples from the simulation. This can be used to pick all clusters in the simulation that belong to the same class as the X-ray cluster sample, \textit{i.e.} could have been observed by ROSAT. This is discussed in Section \ref{sec:realization}. For some applications this is the end of our task. 

\begin{figure}[t!]
\centering
\includegraphics[scale=0.46]{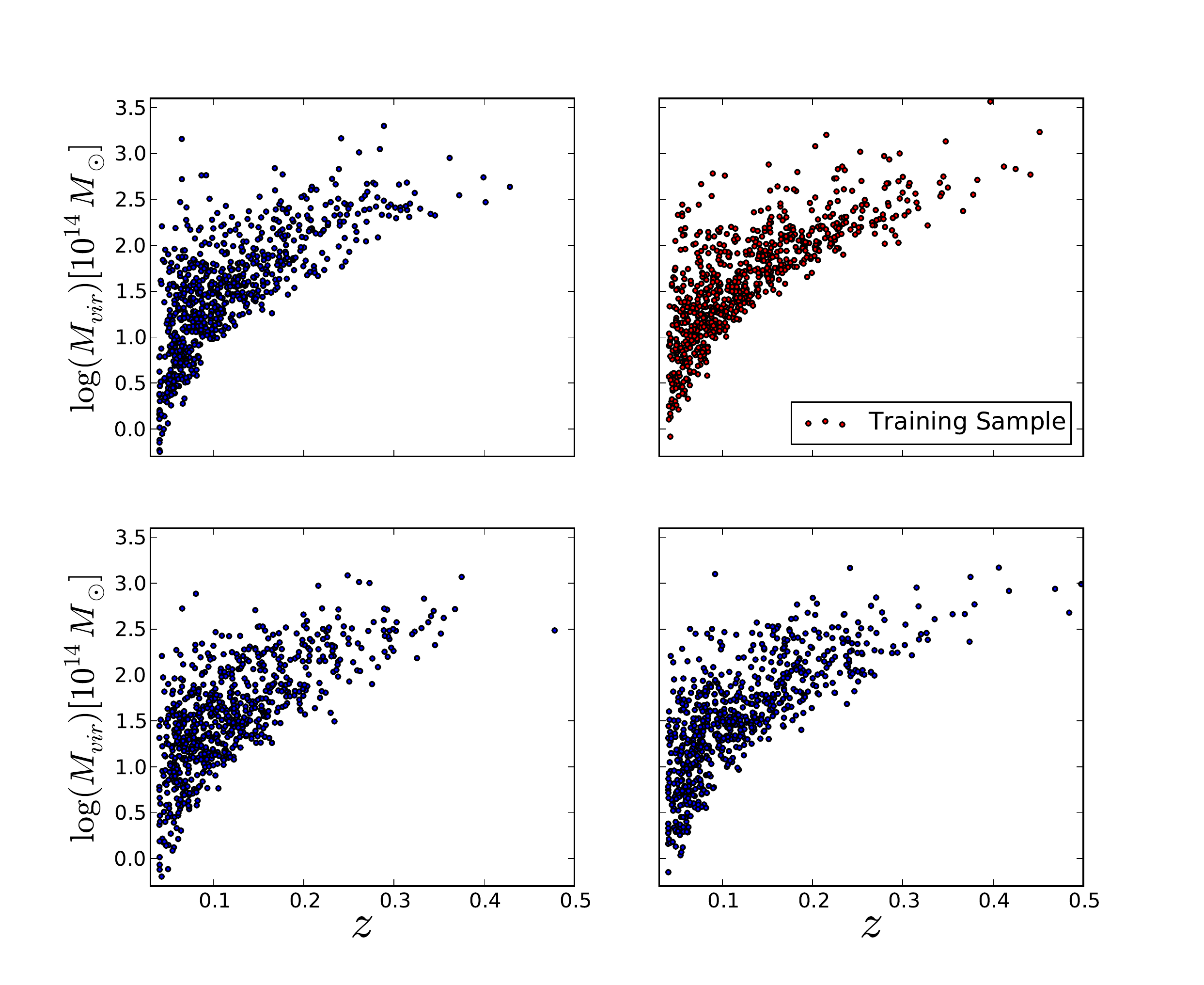}
\caption{Three realizations of the mock cluster catalog using the GMM (\textit{blue}) and the target sample from ROSAT (\textit{top-right}).   }
\label{fig:scatter}
\end{figure}

\begin{figure}[t!]
\centering
\includegraphics[scale=0.4]{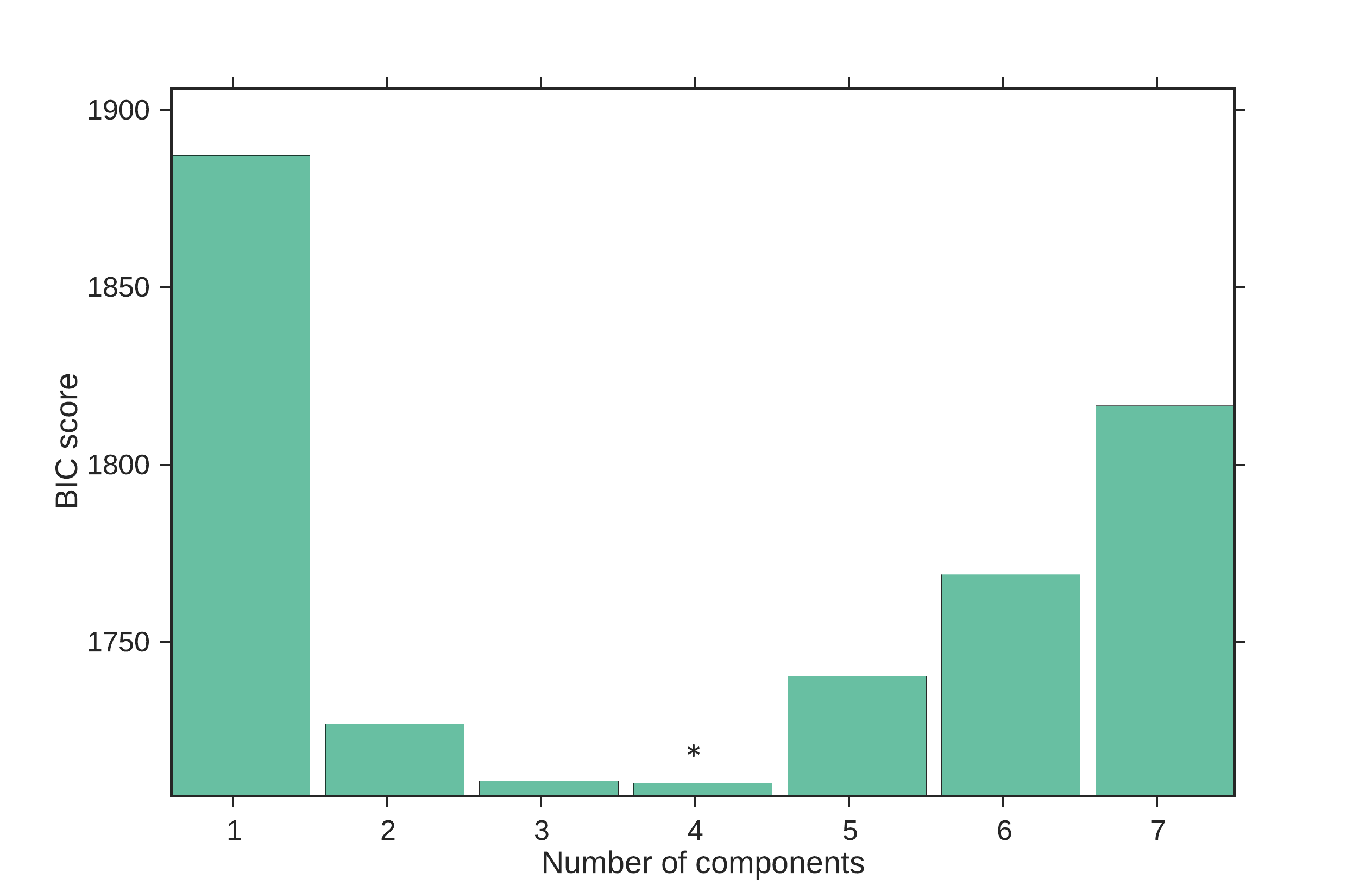}
\caption{The BIC score for the number of components in the Gaussian Mixture Model. a four-component model minimizes the BIC. }
\label{fig:BIC}
\end{figure}

Here though we would like to be more ambitious by trying to pick a sub-sample with the same number of objects as in the ROSAT catalog. And we would like them to statistically have the same probability distribution function in mass and redshift as the ROSAT clusters. To that end, we train a Gaussian mixture model (GMM) to learn the density distribution of the objects in the feature space. GMM's have two free parameter that need to be determined beforehand: the number of mixture components and the form of the covariance matrix to be used. Again we use cross-validation to choose a setup that minimizes the Bayesian information criterion (BIC) \cite{schwarz1978}. The BIC is the logarithm of the maximized likelihood function plus an additional term that penalizes the number of parameters in the model. Minimizing the BIC ensures against overfitting of the data. If there are $n$ data points in the observed data, $\{x\}$, for a model $M$ with $K$ parameters, $\{\theta\}$, the BIC is given by
\begin{equation} \label{eq:BIC}
\mathrm{BIC} = -2\ln{p(x \vert \hat{\theta}}, M) + K \ln{(n)},
\end{equation}
where $p(x \vert \hat{\theta}, M)$ is the likelihood function and $\{\hat{\theta}\}$ are the parameter values that maximize the likelihood function. 
The number of components, $K$ in a GMM are found using the BIC in a simple loop: we let the number of components vary between 0 and $K_{max}$, where $K_{max}$ is chosen sufficiently large (\textit{e.g.} 50 in our analysis) to ensure a full coverage of the parameter space. For every $K$ value we  find the model parameters by maximizing the likelihood function and compute the BIC. Figure \ref{fig:BIC} shows the BIC as a function of number of components, $K$.
The best configuration corresponds to the lowest BIC score with 4 mixture components and a `full' (non-diagonal) covariance matrix.

\subsection{Making Mock Catalogs} \label{sec:realization}

\begin{figure}[t!]
\centering
\includegraphics[scale=0.6]{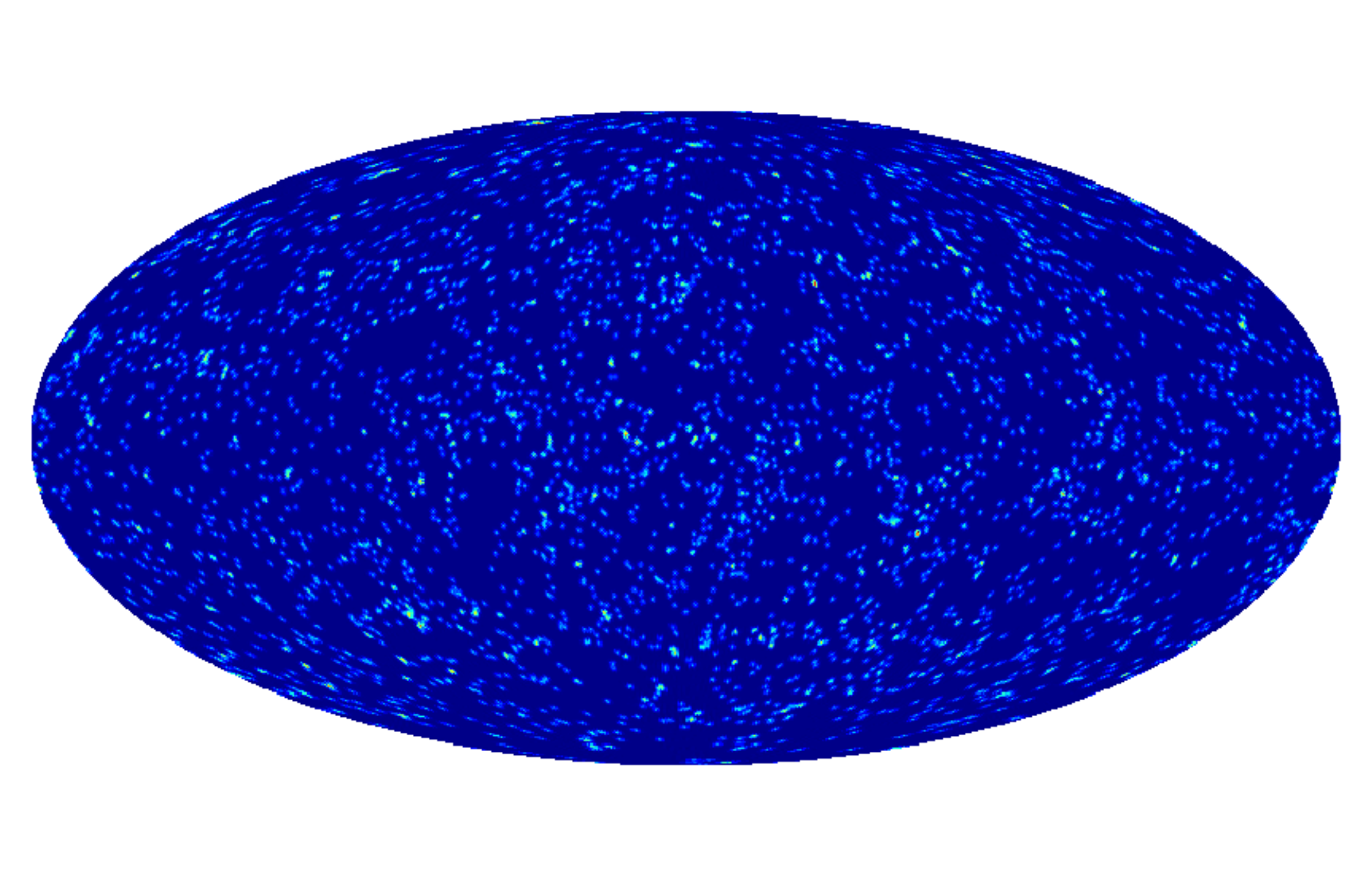}
\caption{An over-density map of a realization of mock catalogs. The map has been smoothed with a 1$^\circ$ Gaussian beam to make the structures more visible. }
\label{fig:simMap}
\end{figure}

The simulated dark-matter halo catalog contains about $10^7$ halos over a wide range of masses and redshifts, much wider than the observed range of these quantities. We use trained classifiers described in the previous section to draw sub-samples of galaxy clusters from the simulated halo catalog that belong to the same class as the observed ROSAT clusters in our training sample. The masses derived from the simulation results are noise-free. But the measured masses of the observed sample contain systematic and stochastic uncertainties\footnote{For a discussion of the magnitudes and sources of these uncertainties see \cite{hajian2013measuring}. }. We impose some uncertainty on the masses of the halos to account for the mass uncertainty in the observed sample. We scale each cluster mass in the simulation by a randomly selected number that is drawn from a Gaussian distribution with a mean of 1 and a standard deviation that matches the $L_\rmn{X}-M$  scatter of 24\% and then determine whether the``noisy" mass, the product of this random number times the true mass, is above the redshift dependent decision boundary. For the decision boundary, we choose one that is generated by a randomly selected training sample as explained in Section \ref{sec:training}.

Thus the procedure for drawing a sample is: shuffle the target sample and choose a random sub-sample containing 80\% of the target clusters; train a one-class SVM classifier with this training sample to determine the decision function; add a Gaussian distributed uncertainty to the masses of clusters; and use the trained classifier to pick those clusters that lie inside the decision boundary. This will generate one realization of clusters belonging to the same class as the observed ROSAT catalog. By repeating this procedure we can generate many realizations based on a single simulation. 
The variance  over the ensemble of these realizations will contain  measurement errors, scaling relation uncertainties and sample variance.  
Using the halos in our simulation, we obtained about 40,000 clusters in each realization. Catalogs drawn this way, \textbf{super-catalogs}, can be used for statistical analysis of the data similar to the example we show in the next section.

\subsubsection{Catalogs with Fixed Densities}

\begin{figure}[t!]
\centering
\includegraphics[scale=0.45]{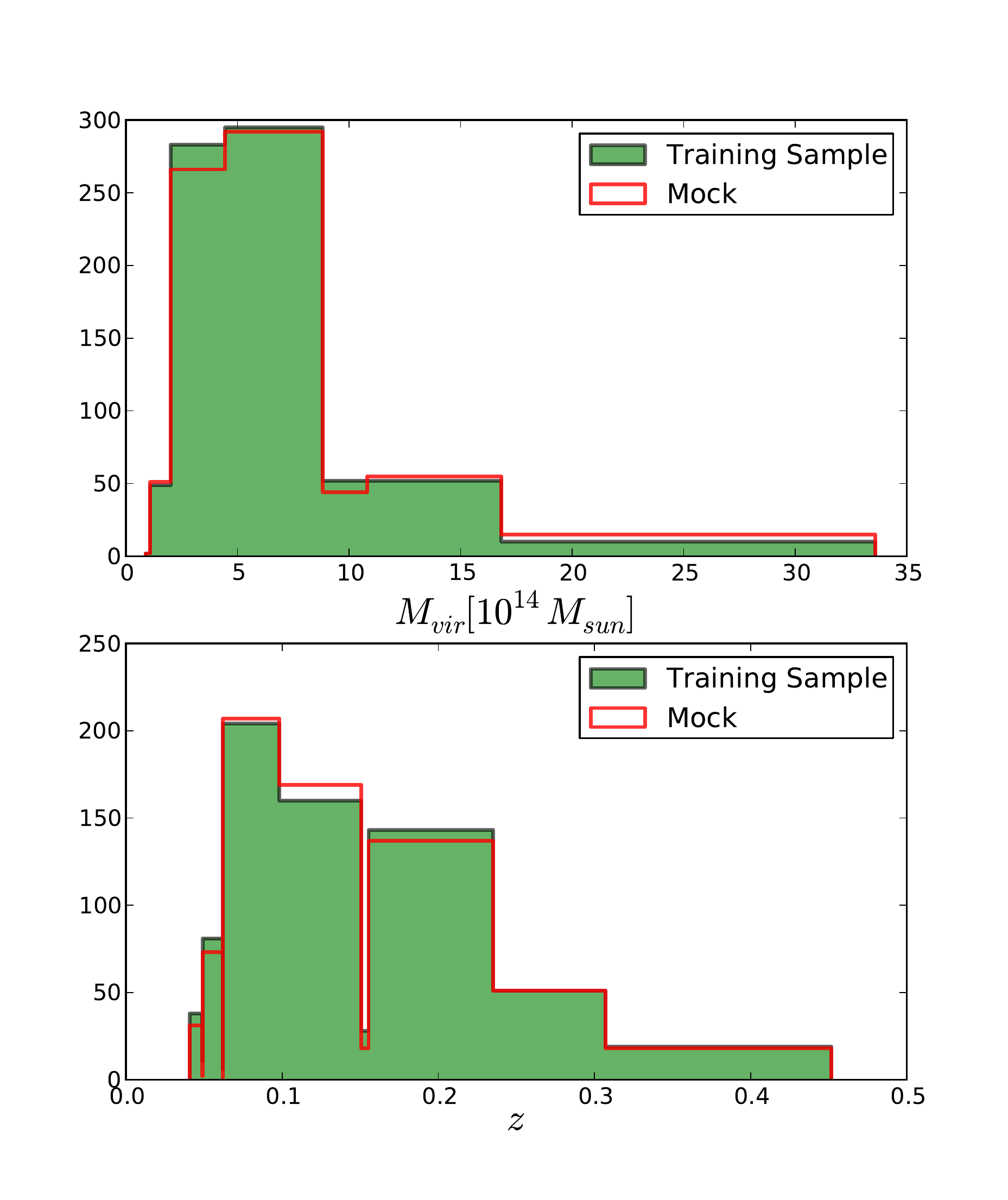}
\caption{Probability distribution function of a single realization of a mock cluster catalog (red) versus that of the target cluster sample from ROSAT (green) for mass (\textit{top}) and redshift (\textit{bottom}) of clusters. The histograms are done using the \textit{Bayesian Blocks} framework of \cite{BayesianBlocks} which uses the data to choose the bin sizes adaptively.}
\label{fig:hist}
\end{figure}

For many astrophysical applications the catalog made by one-class SVM, described in the previous section, is the catalog of interest. However, some problems might need a more constrained catalog. For example we might want the feature space density of the mock catalog to be the same as that of the observed sample. 
To draw cluster samples from the simulation with the same number of objects having statistically the same distribution functions as the target sample, we use the trained GMM classifier from Section \ref{sec:training}. 
The GMM classifier could be run on the full data-set or on the super-catalog made in the previous section.  The former is a straightforward method that can be easily done at the risk of classifying a relatively large number of outliers as the target sample and reducing the purity of the mock catalog.
This is caused by the nature of the distribution of the observed clusters of galaxies in our sample. The sample has a relatively large number of objects to the left side of the GMM components, where the massive clusters reside, and a rather sharp cut-off on the right side. A blind application of the GMM tends to draw samples merely based on the distance from the GMM component centers. This leads to an undesirable classification of outliers as the target class. 
We prefer using the super-catalog as the input set for the GMM to guarantee the purity of the sub-sample. We sample $N$ clusters using a mixture of Gaussian components with the parameters derived from training the GMM on the observed sample data. If the super-catalog had infinitely many objects in it, this would be as simple as drawing $N$ random numbers from a super-position of some multi-dimensional Gaussian random distribution functions. The number of objects in the super-catalog is finite, however it is dense enough to allow us to draw random points in the feature space using the GMM and pick the nearest object to that without replacement. There will be points picked by the GMM that are not allowed by the one-class SVM decision boundary. Those points are discarded and the procedure continues until we obtain $N$ samples.  This will be a single realization of a mock catalog, and can be repeated to generate more catalogs. Here we use $N=820$ to make a mock catalog with the same size as the observed catalog. 
Figure \ref{fig:scatter} shows examples of  mock samples drawn this way. The distribution function of the clusters in the mock catalogs, as shown in Figure \ref{fig:hist}, are very similar to those of the target cluster catalog from ROSAT.

\subsection{Bonus Application: tSZ-Xray Cross-Correlation}
\begin{figure}[t!]
\centering
\includegraphics[scale=0.5]{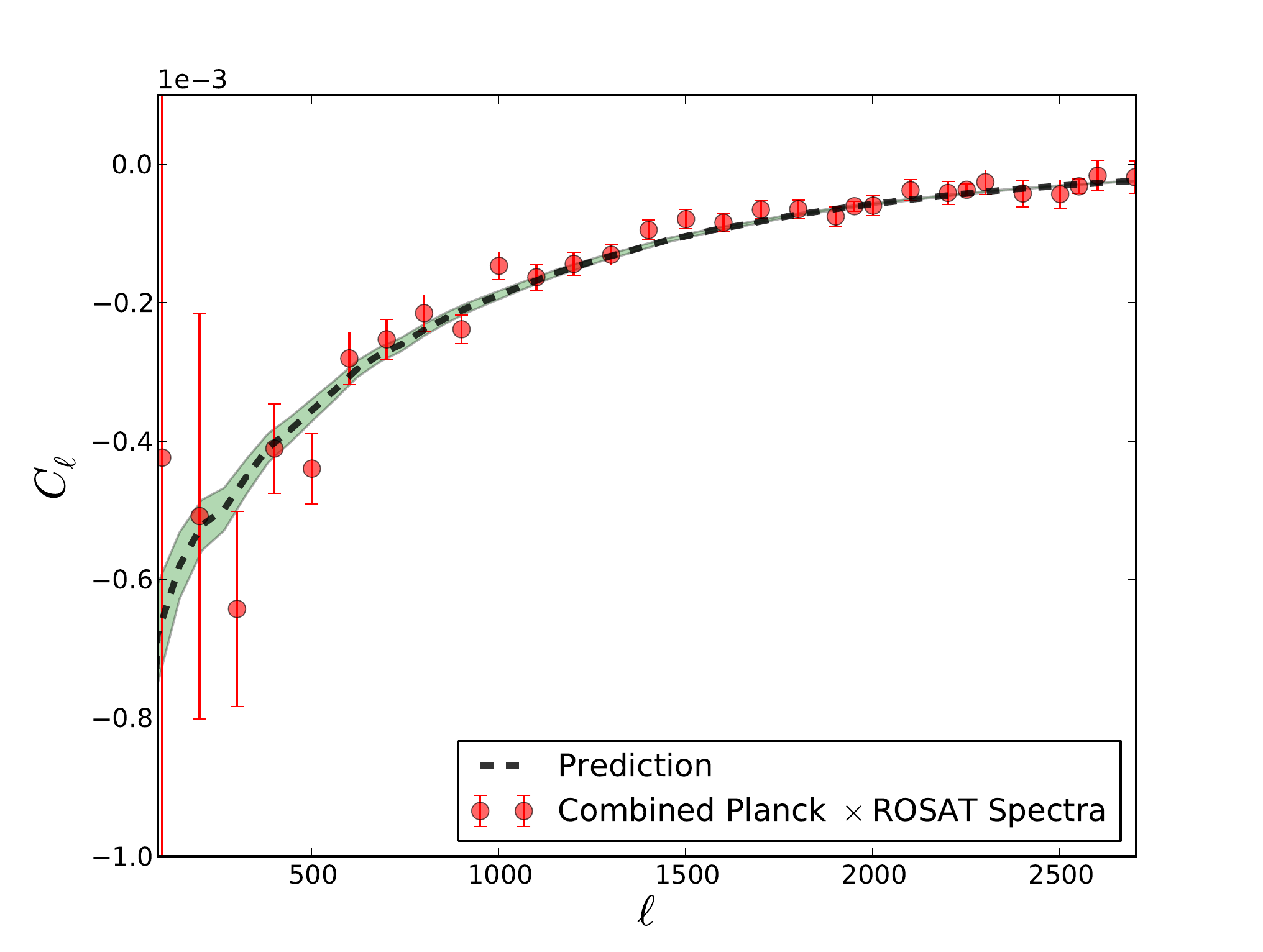}
\caption{A combined noise weighted average of the cross-correlations of Planck with ROSAT clusters (red) from \cite{hajian2013measuring} compared with the average cross-power spectra of the over-density map based on  mock catalogs and a tSZ map based on all halos in the simulation.}
\label{fig:cross1}
\end{figure}

As an example application of the use of the  mock catalogs of Section \ref{sec:realization}, we
simulate cross correlations between Planck thermal Sunya'ev-Zeldovich (tSZ) maps and a number-counts map based on X-ray selected clusters of galaxies from ROSAT.
We use the mock cluster catalog generated in Section \ref{sec:realization} to make a number-counts map and build a normalized number density map based on that. The procedure is the same as the one explained in \cite{hajian2013measuring}: we make the number-counts map ($n(\hat{\mathbf{\theta}})$) first by placing ones in the central pixel locations of clusters. The normalized overdensity map is derived by dividing the number-counts map by its mean value ($\bar{n}$) and subtracting one:
\be 
\delta_{n}(\hat{\mathbf{\theta}}) = \frac{n(\hat{\mathbf{\theta}})}{\bar{n}} - 1. 
\ee
\noindent This, in general, is just a superposition of delta functions convolved with a shape window 
\be \label{eqn:overdensityMap}
\delta_{n}(\hat{\mathbf{\theta}}) = \sum_{i=1}^{N} a_i f(\hat{\mathbf{\theta}}-\hat{\mathbf{\theta}}_i),
\ee
where $f(\hat{\mathbf{\theta}}-\hat{\mathbf{\theta}}_i)$ is the shape we attribute to the clusters in our overdensity map. In the absence of any smoothing (no beam, infinitely small pixels in the map), $f(\hat{\mathbf{\theta}}-\hat{\mathbf{\theta}}_i)$ is a Kronecker delta function. There are various ways of choosing the weights $a_i$. In this analysis we choose uniform constant weights $a_i = a$. Figure \ref{fig:simMap} shows an example of  a full-sky overdensity map made this way based on one realization.  We measure the cross-power spectra of the overdensity maps in eqn. \ref{eqn:overdensityMap} and a tSZ map made from all clusters in the simulation as explained in \cite{pykp}.  We make 200 realizations of mock catalogs and use them to measure the mean cross-power spectrum and the uncertainty on that. The results are shown in Figure \ref{fig:cross1} and are compared with the results of Planck-ROSAT cross-correlations. Red data points show cross-power spectra of Planck and ROSAT clusters from \cite{hajian2013measuring}, dashed black is the average cross-power spectrum from the mock catalogs and the green band around it shows the uncertainty on that.

\section{Discussion and Conclusion}
\label{sec:concl}
We present a set of machine learning classifiers called one-class classifiers to learn the properties of an observed catalog of clusters of galaxies from ROSAT and to pick clusters from mock simulations that resemble the observed ROSAT catalog. The method eliminates  possible inaccuracies due to hand-tuning the selection function in making mock catalogs. It is fast and readily scalable to larger catalogs with a lot more features. We describe a boundary determination method (one-class SVM) for selecting all clusters above a certain redshift-dependent threshold defined by the observed sample.  This method makes no other assumptions about the distribution function of galaxy clusters.  Hence it is useful for constraining cosmology by comparing measurements with predictions from simulations (see \textit{e.g.} \cite{hajian2013measuring}). If we would like our mock catalogs to possess the same distribution function as the observed sample, the density based method (GMM) is the right method to use. The exact choice of the algorithm depends on the astrophysical problem  we would like to address. We show an example application by simulating the measured cross-correlation of  Planck tSZ and ROSAT.

The choice of a discriminative one-class SVM classifier is motivated by the nature of our problem. Unlike in many problems in astrophysics, the target sample of interest here is not a clustered population scattered due to various kinds of uncertainties. Instead, the problem we have here is to find all objects above a redshift-dependent threshold and to reject everything below the threshold as outliers.  it is of particular importance  to make sure all massive objects in the sample are classified as in  the target sample even if they are  far away from the rest of population. This is naturally satisfied with the one-class SVM because high-mass objects are always on the left side of the boundary, hence are never rejected.

An undesirable feature of the SVM classifiers is their hard threshold boundaries. In reality we do not expect the selection function to be perfectly hard-edged. Statistical uncertainties make the real-world selection functions fuzzy and soft. Assuming a sharp selection function can lead to various statistical biases. We soften the SVM decision boundary in two ways to work around this problem: first, we add uncertainties to the masses in the simulation by adding a Gaussian distributed random number to the masses before we make the selection. That is equivalent to making the hard boundary  soft. This is done in the same way as in \cite{hajian2013measuring}. Another fuzziness enters when we choose the boundary.  As it was described in Section \ref{sec:training}, we randomly discard a fraction of the clusters in the training set and obtain the boundary based on the rest of the objects in the sample. This gives us a single boundary surface each time. Figure \ref{fig:learnedSurface} shows the decision boundaries  and also a realization of a fuzzy decision boundary resulting from this procedure. The decision boundaries are tightly squeezed together in this case because the scatter of the data-points near the edge is small. 
The above steps are repeated several times to draw several realizations of the catalog. Therefore in practice we use a family of boundary surfaces and many noise realizations. A random draw of the decision boundary for each realization is a way to include the randomness of the underlying cosmological process of structure formation into account. This effect will always be present due to the stochastic nature of the problem even if the observations are ideal and free from  measurement errors. The second randomization process is designed to take measurement errors and scaling relation uncertainties into account.

The problem we study in this paper only uses two physical properties, namely the mass and the redshift of the objects in the catalog, as features used in classification. Of course, the mass is a derived quantity, but as long as errors are properly included it can be used instead of the more native cluster flux/luminosity and temperature selections. This limited problem might seem simple enough to solve by drawing a line without a need for any sophisticated method. However augmenting the feature space by even one more dimension makes it difficult to solve in this way.
The method here can be used (with minimal modification) to work with many more features. An interesting example is using the positions of the objects in the sky as another feature. This will allow us to make mock catalogs for an inhomogeneous sample with a non-uniform depth coverage in the sky. Another application of adding  objects position as a feature to the classification problem is to obtain a catalog that has clustering properties similar to the observed catalog.  

Using a statistically well-characterized and scalable method to generate the simulated catalog is a major improvement over the manual estimation of a selection function. There are other statistical methods, such as the nearest-neighbor (NN) methods,  that can be used to replicate a distribution. These methods are simple and work well in well-sampled regions of the feature space. But a major drawback in using them for our purpose is their susceptibility  to random patterns in the data. The observed cluster catalog is a single realization of a random process.  The NN methods tend to faithfully keep the observed patterns in the data and hence an accidental void region in the $M-z$ plane will always be present in all catalogs made with the NN algorithms. The one-class SVM algorithm, on the other hand, is a discriminative method solely sensitive to the outermost data-points. Ignoring the internal (random) patterns in the training set, it defines a solid decision function, tagging each object based on its membership to the target class. That is the main feature of the one-class SVM algorithm that makes it properly suited for the application discussed in this paper.

It is often the case for surveys that the observed point process of the galaxies or clusters is considered to be a realization of a continuous selection function, $P(\mathrm{select} \vert q)$, of the set of parameters $\{ q^i , i=1, \cdots, J \}$ being observed. Among the many possibilities for $q^i$ are angular position, redshift, and X-ray flux. For us, we have reduced the parameter space to the derived mass and redshift, both of which have errors associated with them. If the differential number, $dN(q) = n(q) d^J q$, defines the number density in parameter space, $n(q)$, the posterior number density given the selection is 
\begin{equation}
n(q\vert \mathrm{select}) = \frac{P(\mathrm{select} \vert q) n(q)}{P(\mathrm{select})}, 
\end{equation}
where $P(\mathrm{select})$ is an overall normalizing probability, e.g., unity. The observed set of objects defines a specific realization for $n(q\vert \mathrm{select})$, a sum of delta functions. Theoretical mocks of the observations are realizations that we strive to make as realistic as possible, so Monte-Calro analyses of the statistics of the observations can be done. 

Our strategy here is related, but the created mocks are subsets of our all-sky position-space mocks that have learned the sky point process. Complex selection boundaries can be learned this way.
Although we have not explicitly applied flux limits for the application here, it is straightforward to expand to include the flux or position-space restrictions in the training stage.

This example shows the relevance and convenience of unsupervised machine learning algorithms where we do not have enough information to find an exact solution to our astrophysical problems based on first physical principles. This is a promising direction and it will prove even more useful in the near future when a significant growth in the amount of astronomical data will render traditional methods of astrophysical data analysis less effective.

AH would like to thank S. Rezaei for pointing out the possibility of solving such problems with machine learning techniques. 
AH  acknowledges useful discussions with H.R. Maei and A. Mohamad on machine learning parts of this work. We would like to thank J. Vanderplas  for his comments on the manuscript that improved our paper, and  N. Battaglia for  discussions about cluster catalog properties. We would like to thank the anonymous referee whose comments helped us improve our paper. Some of the results in this paper have been derived using \textit{healpy}, the python version of the HEALPix package \citep{healpix}. Research in Canada is supported by NSERC and CIFAR. Some simulations and computations were performed on the GPC supercomputer at the SciNet HPC Consortium. SciNet is funded by the CFI under the auspices of Compute Canada, the Government of Ontario, the Ontario Research Fund Research Excellence; and the University of Toronto.

\appendix

\section{Conversion Relations for Masses} \label{app:mass}
From $M_{500}$ we simultaneously calculate the virial mass, $M_{vir}$, i.e., mass enclosed within the virial radius, $R_{vir}$,  and the concentration
parameter, $c$, using the system of equations
  \bea   \label{eq:system}
    M_{vir} &=& \frac{4\pi}{3}[\Delta_c(z)\rho_c(z)]R_{vir}^3, \\  \nonumber
    M_{500} &=& M_{vir}\frac{m(cR_{500}/R_{vir})}{m(c)},  \\ \nonumber
    c &=& \frac{5.72}{(1+z)^{0.71}}\left( \frac{M_{vir}}{10^{14}h^{-1}M_{\odot}}  \right)^{-0.081},
  \eea
where $m(x) = \ln(1+x) - \frac{x}{1+x}$, $c$ is taken from  \cite{duffy2008dark} based on the N-body simulations with the
WMAP five-year cosmological parameters and $\Delta_c(z)$ is a function of $\Omega_m$ and $\Omega_\Lambda$ \citep{bryan1998statistical},
\be
\Delta_c(z) = 18\pi^2 + 82[\Omega(z) - 1] -39[\Omega(z) - 1]^2,
\ee
with $\Omega(z) = \Omega_m(1+z)^3/E^2(z)$.   For reference, $R_{vir}$  is approximately $2 R_{500}$  \cite{komatsu2011seven}.
We use the concentration parameter to define the scaling radius, namely
 \be
 r_s = \frac{R_{vir}}{c}.
 \ee
We numerically solve the system of equations (\ref{eq:system}) using  the FuncDesigner library for Python\footnote{http://openopt.org/FuncDesignerDoc}.

\bigskip

\bibliographystyle{plain}

\end{document}